%% file: main.tex
\documentclass[a4paper]{spie}  %>>> use this instead for A4 paper
\usepackage{amsmath,amsfonts,amssymb}
\usepackage{graphicx}
\usepackage[colorlinks=true, allcolors=blue]{hyperref}

\title{An efficient approach to global sensitivity analysis and parameter estimation for line gratings}

\author[a,b]{Nando Farchmin}
\author[c,d]{Martin Hammerschmidt}
\author[c,d]{Philipp-Immanuel Schneider}
\author[a]{Matthias Wurm}
\author[a]{Bernd Bodermann}
\author[a]{Markus B\"ar}
\author[a]{Sebastian Heidenreich}
\affil[a]{Physikalisch-Technische Bundesanstalt, Braunschweig and Berlin}
\affil[b]{Technische Universit\"at Berlin, Institute of Mathematics}
\affil[c]{JCMwave GmbH}
\affil[d]{Zuse Institute Berlin}

\authorinfo{E-mail: nando.farchmin@ptb.de}

\begin{document} 
\maketitle

\begin{abstract}
  Scatterometry is a fast, indirect and nondestructive optical method for the
  quality control in the production of lithography masks. Geometry  parameters
  of line gratings are obtained from diffracted light intensities by solving an
  inverse problem. To comply with the upcoming need for improved accuracy and
  precision and thus for the reduction of uncertainties, typically computationally
  expansive forward models have been used. In this paper we use Bayesian
  inversion to estimate parameters from scatterometry measurements of a silicon
  line grating and determine the associated uncertainties. Since the direct
  application of Bayesian inference using Markov-Chain Monte Carlo methods to physics-based partial
  differential equation (PDE) model is
  not feasible due to high computational costs, we use an approximation of the PDE
  forward model based on a polynomial chaos expansion. The expansion provides
  not only a surrogate for the PDE forward model, but also Sobol indices for a
  global sensitivity analysis. Finally, we compare our results for the global
  sensitivity analysis with the uncertainties of estimated parameters. 
% \footnote{The basic format was developed in 1995 by Rick Hermann (SPIE) and
% Ken Hanson (Los Alamos National Lab.).}
\end{abstract}


% ------------------------------------------------------------------------------
\section{INTRODUCTION}
\label{sec:intro}

Scatterometry is an optical scattering technique frequently used for the
characterization of periodic nanostructures on surfaces in semiconductor
industry \cite{HT04}. In contrast to other techniques like electron microscopy,
optical microscopy or atomic force microscopy, scatterometry is a
non-destructive and indirect method. In particular, geometry parameters of
interest are determined by measuring diffraction patterns and solving an
inverse problem. A basic requirement for the success of the estimation of
parameters is that the underlying model is sensitive to the parameters of
interest. The more sensitive a systems dependence on a certain parameter, the
easier is the reconstruction of that parameter.  On the other hand, when
designing numerical models to simulate experiments, it is often unclear which
parameters are necessary for the model, often resulting in a large amount of
parameters used for the model even if some of them have little or no influence
on the system.  For both reasons, a sensitivity analysis is often useful. A
sensitivity analysis gives a priori information about the influence of input
parameters on the output.

Most often, variance based sensitivity indices are computed by using
Monte-Carlo methods, which is computationally demanding \cite{HS96,SAA+10}.
Hence, often only the sensitivities of a few, presumably most important parameters are
considered.  Here we present an expansion into polynomials that yields an
algebraic approach to characterize the global sensitivities for all parameters
with less computational cost \cite{Sud08}. 

%Compared to the widely used least squares method, the application of a maximum
%likelihood estimation (MLE) method achieved improved reconstruction results for
%geometry parameters in the recent years.  Moreover, MLE methods allow for the
%modelling of additional error parameters to handle measurement noise. However,
%MLE is not applicable to combined measurements, as often obtained by
%experiments. 
%Bayesian inference on the other hand is a statistical sampling
%method that requires a large number of function evaluations which is not
%feasible for the expansive computations of today's scattering forward models.
%In this paper we show how a surrogate model obtained through a polynomial chaos
%expansion allows for implementing a Bayesian approach to the inverse problem.  

\begin{figure} [ht]
  \begin{center}
    \begin{tabular}{c} %% tabular useful for creating an array of images 
      \includegraphics[width=.8\linewidth]{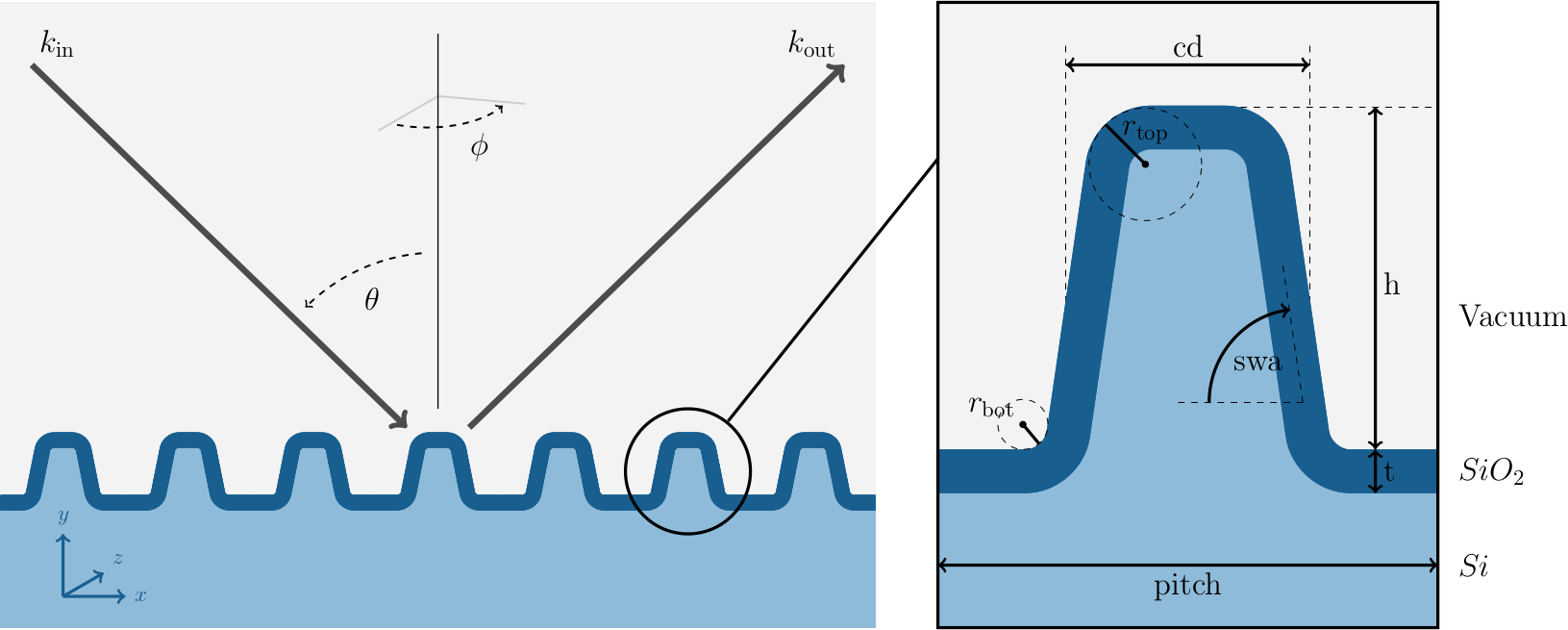}
    \end{tabular}
  \end{center}
  \caption{\label{fig:scat_setup} 
    Cross section of the photomask with description of the stochastic
    parameters. The dimensional parameter vector is given by
    $\xi=(h,\mathrm{cd},\mathrm{swa},t,r_\mathrm{top},r_\mathrm{bot})$.  The
    pitch, i.e. the length of periodicity is fixed to $50\,\mathrm{nm}$.
    }
\end{figure} 

In this paper, we determine the geometry parameters of a photomask that consists
of multilayered, periodic, straight absorber lines of two optically different
materials. The period of the line structure (pitch) is $50\,\mathrm{nm}$ and
the geometry parameters of interest are the height of the line $h$, the width
at the middle of the line (critical dimension) $\mathrm{CD}$, the sidewall
angle $\mathrm{SWA}$, the silicon oxide layer thickness $t$ and the radii of
the rounding at the top and bottom corners of the line $r_\mathrm{top}$ and
$r_\mathrm{bot}$, respectively. A cross section of the geometry for one period
of the structure is depicted in Fig.~\ref{fig:scat_setup}. The photomask was
illuminated by a light beam of wavelength $\lambda=266\,\mathrm{nm}$ for
different angles of incidence $\theta=3^\circ,5^\circ,\dots,87^\circ$ for
perpendicular ($\phi=0^\circ$) and parallel ($\phi=90^\circ$) orientation of
the beam with respect to the grating structure as well as S and P polarization.

In the next sections, we will proceed as follows. First, we introduce the
forward model of the problem, followed by the global sensitivity analysis and
Bayesian inversion based on a polynomial chaos expansion.  Second, we execute
the global sensitivity analysis and estimate the posterior distribution from
measurement data. Finally, we compare both results. 

% ------------------------------------------------------------------------------
\section{FORWARD MODEL}
\label{sec:forward_model}

In principle, the propagation of electromagnetic waves is described by
Maxwell's equations, but for our simple grating geometry Maxwell's equations
reduce to a single second order partial differential equation \cite{HWS+17}, 
\begin{align}
  \label{eqn:Helmholtz}
  \operatorname{\nabla\times} \mu^{-1} \operatorname{\nabla\times} E 
  + \omega^2 \varepsilon \, E = 0.
\end{align}
Here, $\varepsilon$ and $\mu$ are the permitivity and permeability and
$\omega$ is the frequency of the incoming beam. The boundary conditions are
chosen to be transparent in horizontal and periodic in lateral direction.  The
resulting boundary value problem is solved by a finite element method (FEM).
For computations we used the
JCMsuite~\footnote{\url{https://jcmwave.com/jcmsuite}} software package.

The forward model is given by a map of geometry parameters onto S and P
polarization of first order intensities of the scattered light. The
parameters $\xi$ used for modelling the grating geometry are depicted in
Fig.~\ref{fig:scat_setup}. The forward model is represented by the function
$f^*\colon \Omega\to\mathbb{R}^d$ such that the parameters
$\xi\in\Omega\subset\mathbb{R}^M$ are mapped to diffracted efficiencies for a
set of azimuthal angles, incidence angles and polarizations.

To obtain a fast evaluation of the surrogate, the
function $f^*$ is expanded into an orthonormal polynomial basis
$\{\Phi\alpha\}_{\alpha\in\Lambda}\subset
L^2(\Omega;\varrho)$\cite{Wie38,CM47,GS90}
\begin{align}
  \label{eqn:pce}
  f^*(\xi) \approx f(\xi) = \sum_{\alpha\in\Lambda} f_\alpha \Phi_\alpha(\xi)
  \qquad\text{with}\qquad
  f_\alpha = \int_{\Omega}^{}f(\xi)\, \Phi_\alpha(\xi) \,\mathrm{d} \varrho(\xi).
\end{align}
The finite set $\Lambda\subset\mathbb{N}_0^M$ is a set of multiindices and
$\varrho$ denotes the multivariate parameter density for the parameters $\xi$.
With this surrogate the evaluation of the model in different parameter
realizations is equivalent to the evaluation of polynomials. Additionally, we
remark that this is a nonintrusive method and hence the solver used for the
deterministic Helmholtz equation \eqref{eqn:Helmholtz} needs no adaptation
\cite{Sud08}.

In our approach, the experimental data $y\in\mathbb{R}^d$ are modelled
with the error $y_j = f_j(\xi) + \varepsilon_j$, $j=1,\dots,d$ where
$\varepsilon_j\sim\mathcal{N}(0,\sigma_j)$ describes a normal distributed noise
with zero mean, standard deviation and error parameter $b$,
\begin{align}
  \label{eqn:noise_variance}
  \sigma_j(b) = b\, y_j,
  \qquad \text{for } b > 0.
\end{align}
The inverse problem in scatterometry is
defined by the determination of geometry parameter values $\xi$ and the error
parameter (hyper parameter) $b$ from measured efficiencies $y$.

% ------------------------------------------------------------------------------
\section{GLOBAL SENSITIVITY ANALYSIS}
\label{sec:sensitivity}

Sensitivity analysis is a broadly used tool to identify the influence of
uncertain input parameters upon the output of a physical system or model.
Local methods for sensitivity analysis utilize partial derivatives of the
output of the system with respect to the various uncertain input parameters to
obtain the local parameter dependence of the system \cite{SA10}. However, since
local sensitivity analysis does not cover the hole input space, only small
perturbations can be observed. Global variance-based sensitivity analysis on
the other hand decomposes the total system variance $\operatorname{Var}[f^*]$
over the complete parameter space into parts attributing to input parameters
and combinations thereof \cite{Sob93,Sob01}. Among the vast collection of
variance-based methods for sensitivity analyses, Sobol indices are a common and
widely spread method to characterize parameter sensitivities. The map $f^*$
from above with expectation $\mathbb{E}[f^*]$ and variance
$\operatorname{Var}[f^*]$ can be decomposed as \cite{Sob93}
\begin{align}
  \label{eqn:sobolDecomposition}
  f^*(\xi) 
  = S_0 + \sum_{1\leq s\leq M}^{}\ \sum_{j_1<\dots<j_s\in\mathcal{M}}
  S_{j_1,\dots,j_s}(\xi_{j_1},\dots,\xi_{j_s})
  \qquad\text{for }\mathcal{M}=\{1,\dots,M\},
\end{align}
into functions $S_{j_1,\dots,j_s}$ depending exactly on the parameters
$\xi_{j_1},\dots,\xi_{j_s}$. Inserting this decomposition into the computation
of the variance of the map $f^*$ then yields the Sobol indices, i.e. 
\begin{align}
  \label{eqn:sobolIndices}
  \operatorname{Sob}_{j_1,\dots,j_s} 
  = \frac{\operatorname{Var}[S_{j_1,\dots,j_s}]}{\operatorname{Var}[f^*]}.
\end{align}
Computing the variances in the equation above requires high dimensional
integration. These integration is often calculated by Monte-Carlo methods that
require many expensive function evaluations.

It was previously shown \cite{Sud08}, that the uniqueness of the PC expansion
and the uniqueness of the Sobol decomposition \eqref{eqn:sobolDecomposition},
gives the algebraic equivalence
\begin{align}
  \label{eqn:sobolIndexWithPCE}
  \operatorname{Sob}_{j_1,\dots,j_s} 
  \approx \frac{\sum_{\alpha\in\Lambda_{j_1,\dots,j_s}} f_\alpha^2}
  {\sum_{\alpha\in\Lambda\setminus\{0\}}f_\alpha^2}.
\end{align}
Here $\Lambda_{j_1,\dots,j_s}\subset\Lambda$ is the set of multiindices that
differ from zero in exactly the components $j_1,\dots,j_s$. Note,
that the "$\approx$" in \eqref{eqn:sobolIndexWithPCE} is a result of
truncating to finitely many terms in the PC expansion.  If the complete PC
expansion is used, equality holds in \eqref{eqn:sobolIndexWithPCE}.

% ------------------------------------------------------------------------------
\section{BAYESIAN APPROACH}
\label{sec:bayes}

The Bayesian approach provides a statistical method to solve the inverse
problem. Following Bayes' theorem, the posterior density is given by
\begin{align}
  \label{eqn:posterior}
  \pi(\hat\xi;y) 
  = \frac{\mathcal{L}(\hat\xi;y) \pi_0(\hat\xi)}
  {\int\mathcal{L}(\hat\xi;y)\pi_0(\hat\xi)\,\mathrm{d}\hat\xi},
\end{align}
where the prior density $\pi_0$ describes prior knowledge and the likelihood
function $\mathcal{L}$ contains the information obtained from the measurement.
The vector $\hat\xi$ consists of geometry parameters $\xi$ and the noise
parameter, i.e. $\hat\xi = (\xi,b)$. Assuming normal distributed measurement
errors, we choose the likelihood function \cite{HGB15}
\begin{align}
  \label{eqn:likelihood}
  \mathcal{L}(\hat\xi;y)
  = \prod_{j=1}^d \frac{1}{\sqrt{2\pi}\sigma_j(b)}
  \exp\left( -\frac{(f_j(\xi)-y_j)^2}{2\sigma_j^2(b)} \right).
\end{align}
In the Bayesian framework, the distributions of parameters are in general
determined by Markov Chain Monte Carlo (MCMC) sampling where for every sampling
step, the forward model has to be evaluated.  Normally, this means that the
Helmholtz equation has to be solved which makes MCMC sampling impractical due to
the large number of required sampling steps. Since the surrogate only requires
evaluations of polynomials, the Bayesian approach becomes practical for
scatterometry measurement evaluations\cite{HGB18}. 

% TODO rel.std in % or in decimal?
\input{table.tex}

For Bayesian inversion, we have to choose a prior distribution for the
parameters, calculate the likelihood function and determine the
corresponding posterior distribution. The posterior distribution contains the
desired parameter values and their associated uncertainties. When two or more
measurement results from different measurement sets $y^{(1)},\dots,y^{(K)}$ are
combined, the posterior distribution of the first measurement can be used as
the prior distribution for the evaluation of the second measurement, i.e.
\begin{align}
  \label{eqn:combined_measurements}
  \pi(\hat\xi;y^{(K)},y^{(K-1)},\dots,y^{(1)}) 
  = \frac{\pi_0(\hat\xi) \prod_{k=1}^K\mathcal{L}(\hat\xi;y^{(k)}) }
  {\int\pi_0(\hat\xi) \prod_{k=1}^K\mathcal{L}(\hat\xi;y^{(k)})
  \,\mathrm{d}\hat\xi}.
\end{align}
Note that the model function $f$ in the likelihood function is in general
different for different measurement setups.
% For example, the grating geometry of the photomask can be measured by an
% atomic force microscope (AFM) and by scatterometry.

\begin{figure} [ht]
  \begin{center}
    \begin{tabular}{c} %% tabular useful for creating an array of images 
      \includegraphics[width=.9\linewidth]{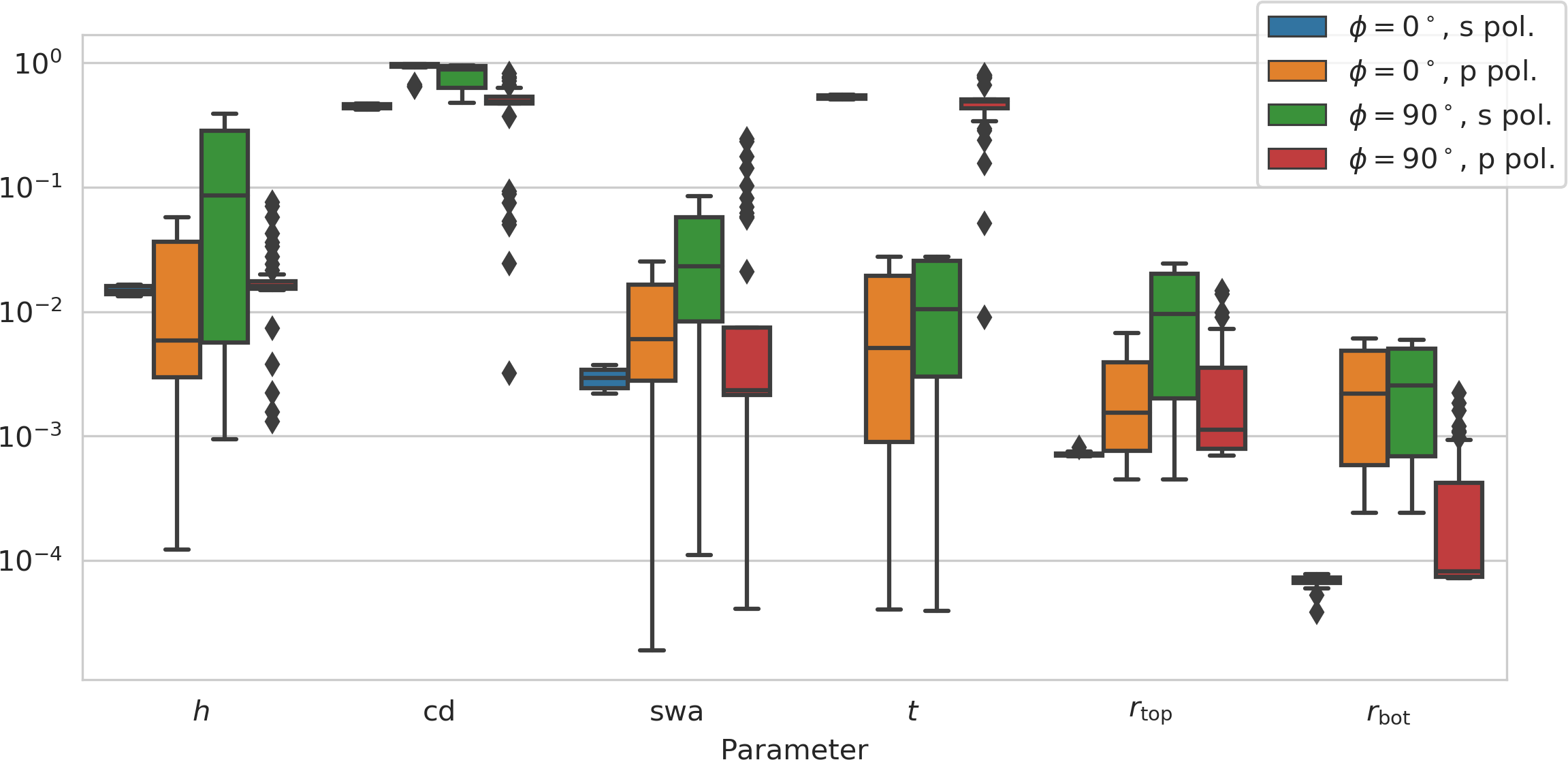}
    \end{tabular}
  \end{center}
  \caption{\label{fig:boxplot_sobol} 
   Boxplots of Sobol indices for all geometry parameters, azimutal angles and
   polarization. Each boxplot includes Sobol indices for the whole range of
   angles of incidence.}
\end{figure} 

%Even though the surrogate model can be evaluated in parameter values outside of
%the original approximation domain, a valid approximation of the surrogate can
%only be guaranteed within the original parameter intervals. However, for the
%reconstruction, we decided to search for parameter values in an enlarged
%domain. If the posterior distribution suggests contributions outside of the
%original set, it can be read as an indicator to change the approximation domain
%for the surrogate and rerun the reconstruction.  
%The reconstruction domains for
%the parameters are displayed in Table~\ref{tab:parameter}.  In our study we
%used uniform distributions for the prior for all parameters including the error
%parameter $b$.

% ------------------------------------------------------------------------------
\section{RESULTS}
\label{sec:results}
For the geometry presented above, we calculated the Sobol indices
Eq.(\ref{eqn:sobolIndices}) for all parameters as depicted in Fig.
~\ref{fig:boxplot_sobol}. 
% Since Sobol indices represent the contributions of (a combination of)
% parameters to the total variance, we can infer the sensitivity of those
% parameters. If the contribution to the total variance is large (Sobol index
% close to $1$), the system is sensitive to the changes in these parameters.  In
% Fig.~\ref{fig:boxplot_sobol} the Sobol indices are depicted as boxplots. 
Boxplots for perpendicular and parallel beam incidence as well as S and P
polarization are shown, respectively, over $43$ angles of incidence $\theta$.
The sensitivity parameter correlation is very small and is therefore omitted
here. First of all, we note that the height, critical dimension and oxide layer
thickness make up most of the total variance. The sensitivity depends also on
the polarization and the angle of incidence. For example, the oxide layer
thickness $t$ is most sensitive with respect to the S polarization for a
perpendicular beam orientation ($\phi=0^\circ$) and the sidewall angle depends
highly on the angle of incidence (P polarization, parallel to the beam).

With this, we expect that the reconstruction of all parameters is feasible.
In particular, the oxide layer thickness and critical dimension should be
possible to determine precisely due to their high sensitivity. 

Next, we apply Bayesian inversion on scatterometry measurements to estimate
geometry parameters of the line grating. More details of the measurement setup
are described in \cite{WBBR11,HWS+17}. For Bayesian inversion it is necessary
to chose prior distributions. In our investigations we have chosen uniform
priors on the domains given in Table~\ref{tab:parameter}.  To obtain the
posterior distribution, we sampled with a MCMC random walk Metropolis algorithm
using the surrogate. 
% At each step, the surrogate of the forward model has to be evaluated, since
% typical sampling sizes of more than $10^5$ samples are impractical with the
% FEM forward model. 
We have chosen a sampling size of $10^6$ samples and a burn in phase of $10^4$
samples. For diagnostics, we applied the Gelman-Rubin criterion \cite{GR92}, to
assure that generated samples are independent.
%\begin{figure} [ht]
%  \begin{center}
%    \begin{tabular}{c} %% tabular useful for creating an array of images 
%      \includegraphics[width=.9\linewidth]{images/marginal_densities.png}
 %   \end{tabular}
 % \end{center}
 % \caption{\label{fig:marginal_densities} 
 % Marginal 1D and 2D densities for the posterior of the stochastic parameters.
 % For the 1D densities, the mean (solid line) and the standard deviation 
 % (dashed line) are depicted as well.
 % }
%\end{figure} 
%In Fig.~\ref{fig:marginal_densities} the posterior (marginal) densities for all
%$6$ stochastic parameters are shown.
We calculated the marginal posterior densities for all $6$ stochastic parameters. All posterior densities are characterized by sharp peaks
with mean and standard deviation similar to the previous publication
\cite{HWS+17}. 
The mean and standard deviation for each parameter including the hyperparameter
(error parameter) $b$ are shown in Table~\ref{tab:parameter}.  Since the domain
sizes of the parameters vary due to their geometrical meaning, we introduce the
relative standard derivation (std). The relative std is the std divided by half
the width of the parameter domain:
\begin{align}
  \label{eqn:relstd}
  \mathrm{rel. std}(\eta) = \frac{2\ \mathrm{std}(\eta)}{\beta-\alpha}
  \qquad\text{where }\eta \in [\alpha,\beta].
\end{align}
The relative std shows how the posterior distribution is spread within the
domain.  For example, if the domain for the critical dimension is $[22,28]
\,\mathrm{nm}$, and the std is $0.3\,\mathrm{nm}$ then the relative std is
$\mathrm{rel. std}=0.099$, i.e. the posterior distribution is concentrated in
about $10\%$ of the original domain.  This way we can deduce how wide the
parameter distributions are spread across the reconstruction domains.  The
relative std in Table~\ref{tab:parameter} shows that the smallest
reconstruction uncertainties are obtained for the critical dimension with
relative std about $10\%$, followed by the oxide layer thickness with $18\%$
relative std. The height has a relative std of about $31\%$. The posterior
densities of the sidewall angle and the corner rounding are slightly wider
distributed at about $45\%$ relative std.  This goes in line with the global
sensitivity analysis (see Fig.~\ref{fig:boxplot_sobol}).

The results for the error parameter $b$ depicted in
Table~\ref{tab:parameter} show that the relative measurement uncertainty is
approximately $1\%$. 

\begin{figure} [ht]
  \begin{center}
    \begin{tabular}{c} %% tabular useful for creating an array of images 
      \includegraphics[width=.7\linewidth]{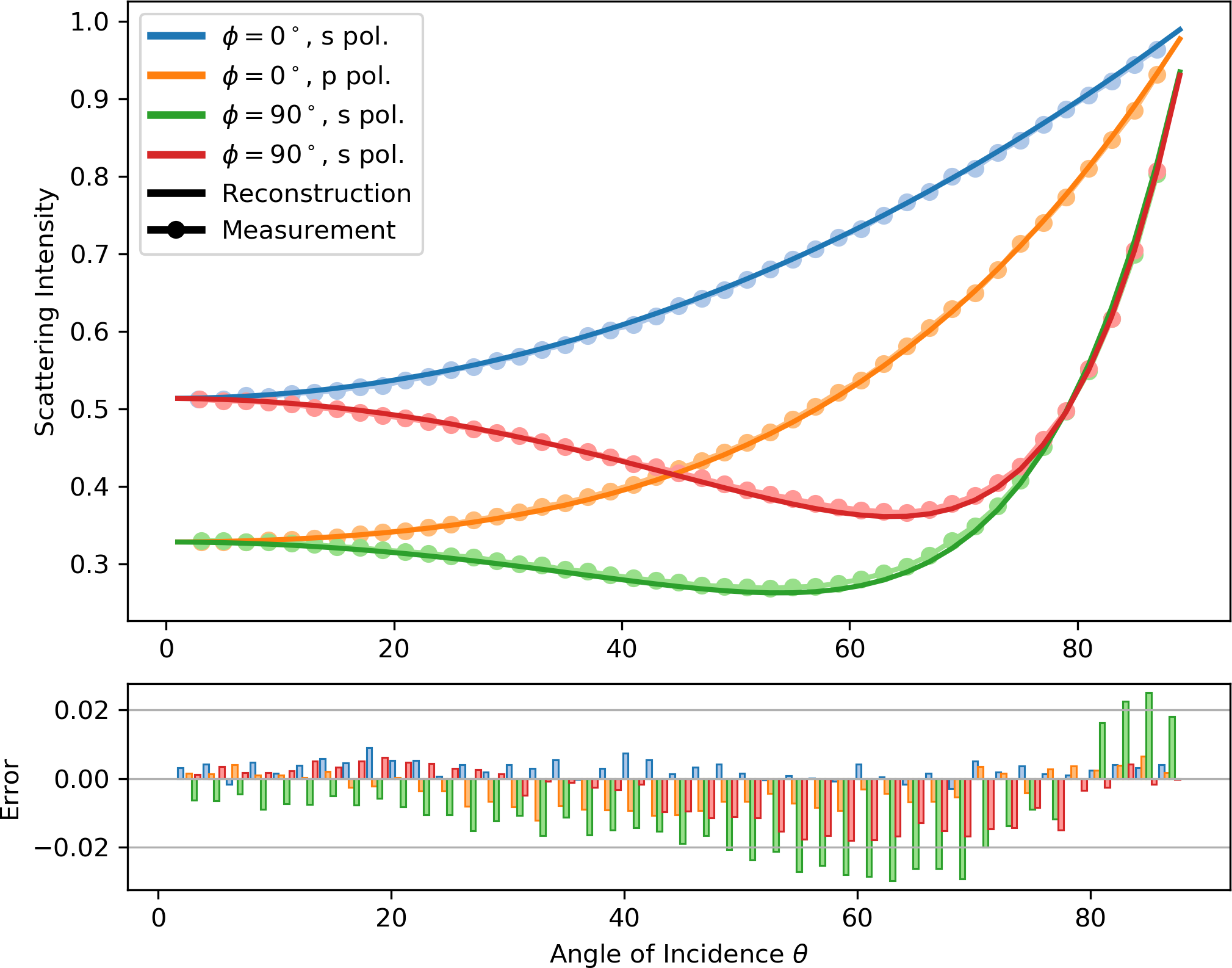}
    \end{tabular}
  \end{center}
  \caption{\label{fig:reconstruction} 
    Scattered intensities for the two polarizations and different azimuthal
    angles. Compared are the measurements of the scatterometry experiment
    and the simulation of the PC surrogate for the mean values of the 
    parameter reconstruction. The bottom graph shows the pointwise deviation.
    }
\end{figure}

Finally, Fig.~\ref{fig:reconstruction} displays a comparison between the
measurement data and the evaluation of our surrogate model using reconstructed
geometry parameters.  The pointwise relative deviation of the approximation
from the measurements data is $2\%$ and lower.  In \cite{HWS+17} the
measurement data were evaluated by a Maximum Posterior Approach (MPA).  The MPA
searches the global maximum of the posterior and uncertainties are determined
by the local covariance matrix. The difference here is that we calculated the
whole posterior distribution. This has the advantage that even for multiple
peaked and non-Gausian posterior distributions this scheme gives reliable
uncertainty estimations.  The results obtained in \cite{HWS+17} are consistent
to our findings. There are only slight differences. For example, the marginal
distribution for the height $h$ is broad (non-Gausian) yielding larger
uncertainties. Similarly, the mean values for $r_{top}$ and $r_{bottom}$ are
slightly shifted due to the asymmetry of the marginal posterior (non-Gausian).
The deviation between the forward model values and the measurement data of
$2\%$ is comparable with that found in \cite{HWS+17}.

%Since the
%pointwise error is comparable to the results in \cite{HHWS+17}, the polynomial
%chaos approach seems to yield at least the same approximation result as other
%methods while being magnitudes faster. Moreover, since the forward model used
%throughout this work was essentially the same as in \cite{HHWS+17}, it is
%possible that the $2\%$ error stems from the accuracy of the underlying forward
%model and does not mirror the approximation capabilities of the polynomial
%chaos approach.

% ------------------------------------------------------------------------------
\section{SUMMARY}
\label{sec:summary}

% TODO last comment from Philipp: numbers for estimated time MCMC with FEM,
% time for PC surrogate/evaluation
% ---- 
% NOTE Evaluation of forward model takes approximately 97s, PC requires 10^4
% evaluations total, each other MCMC version (comparable) 10^6, creating
% surrogate takes 30s, evaluation of surrogate takes 0.05s

In this paper we applied a polynomial chaos expansion as a surrogate for the
forward model in scatterometry. This approach enables us to perform a global
sensitivity analysis at low computational costs.  Moreover, since the surrogate
only requires the evaluation of polynomials instead of solving the Helmholtz
equation, it was feasible to use a full Bayesian approach to determine the
posterior distribution for all geometry parameters. To generate samples from
the posterior distribution, we employed a MCMC Metropolis random work sampling
method and checked the overall independence of the samples obtained by the
Gelman-Rubin criterion. The reconstruction results obtained by the surrogate
model compared to those obtained by a Maximum Posterior estimate with a
Gauss-Newton like method \cite{HWS+17} are consistent and are in line with the
predictions from our global sensitivity analysis. We finally conclude that a
Bayesian approach based on the polynomial chaos surrogate gives accurate and
reliable estimations for silicon line grating parameters and uncertainties.

%Finally, we want to emphasize that the polynomial
%chaos expansion of the forward model with respect to the stochastic parameters
%is non-intrusive, hence the underlying (FEM) solver has not to be adjusted to
%handle stochastic uncertainties.

% ------------------------------------------------------------------------------
% References
\nocite{*}
\bibliography{report} % bibliography data in report.bib
\bibliographystyle{spiebib} % makes bibtex use spiebib.bst

\end{document}

%% file: table.tex
\begin{table}[ht]
  \caption{ \label{tab:parameter}
  Estimations of parameters and uncertainties obtained from the mean value
  (mean), the standard deviation (std) and relative standard deviation
  (rel.std) of the posterior distribution. The domain indicates the support of
  the prior distribution chosen.
  }
  \begin{center}       
    \begin{tabular}{|p{2cm}p{2.5cm}p{2cm}p{2cm}p{2cm}|}
      \hline
      \rule[-1ex]{0pt}{3.5ex}  
      parameter & domain & mean & std & rel.std \\
      \hline
      \rule[-1ex]{0pt}{3.5ex}  
      $h$ 
      & $[43.0 , 53.0]\,\mathrm{nm}$ 
      & $48.35\,\mathrm{nm}$ 
      & $1.56\,\mathrm{nm}$ 
      & $0.3112$ \\
      \rule[-1ex]{0pt}{3.5ex}  
      $\mathrm{cd}$ 
      & $[22.0 , 28.0]\,\mathrm{nm}$ 
      & $25.48\,\mathrm{nm}$ 
      & $0.30\,\mathrm{nm}$ 
      & $ 0.099$ \\
      \rule[-1ex]{0pt}{3.5ex}  
      $\mathrm{swa}$ 
      & $[84.0 , 90.0]^\circ$ 
      & $86.87^\circ$ 
      & $1.33^\circ$ 
      & $0.4424$ \\
      \rule[-1ex]{0pt}{3.5ex}  
      $t$ 
      & $[ 4.0 ,  6.0]\,\mathrm{nm}$ 
      & $ 4.96\,\mathrm{nm}$ 
      & $0.18\,\mathrm{nm}$ 
      & $0.1756$ \\
      \rule[-1ex]{0pt}{3.5ex}  
      $r_\mathrm{top}$ 
      & $[ 8.0, 13.0]\,\mathrm{nm}$ 
      & $10.65\,\mathrm{nm}$ 
      & $1.15\,\mathrm{nm}$ 
      & $0.4601$ \\
      \rule[-1ex]{0pt}{3.5ex}  
      $r_\mathrm{bot}$ 
      & $[ 3.0 , 7.0]\,\mathrm{nm}$  
      & $ 4.89\,\mathrm{nm}$  
      & $0.90\,\mathrm{nm}$  
      & $0.4495$ \\
      \hline
      \rule[-1ex]{0pt}{3.5ex}  
      $b$
      & $[ 0.0 ,  0.1]$
      & $0.01$
      & $0.0021$
      & $0.0849$ \\
      \hline
    \end{tabular}
  \end{center}
\end{table}